\documentstyle[11pt,aaspp4]{article}
\begin{document}

\title{Neutrino afterglow from Gamma-Ray Bursts: $\sim 10^{18}$ eV}
\author{Eli Waxman$^1$ and John N. Bahcall$^{1,2}$}
\affil{$^1$Department of Condensed-Matter Physics, Weizmann Institute,
Rehovot 76100, Israel}
\affil{$^2$Institute for Advanced Study, Princeton, NJ 08540}

\begin{abstract}
We show that
a significant fraction of the energy of a  $\gamma$-ray burst (GRB) is
probably converted to a burst of $10^{17}$--$10^{19}$~eV neutrinos
and multiple GeV $\gamma$'s that follow
the main GRB by $\sim  10$ s.
If GRB's accelerate protons to $\sim 10^{20}$ eV,
a suggestion that recently gained support from observations of GRB afterglow,
then both the neutrinos and the gammas may be detectable.
\end{abstract}

\section{Introduction}

The widely accepted interpretation of the
phenomenology of $\gamma$-ray bursts (GRB's) is that the
observable effects are due to the dissipation of the kinetic energy
of a relativistically expanding fireball whose primal cause is not yet known
[see M\'esz\'aros (1995) and Piran (1996) for reviews].
The physical conditions in the dissipation region imply that protons can
be Fermi accelerated to energies $> 10^{20}$ eV (Waxman 1995a; Vietri
1995; see Waxman 1999 for recent review).

Adopting the conventional fireball picture,
we showed previously
that the prediction of an accompanying  burst  of $\sim10^{14}{\rm eV}$
neutrinos is a natural consequence (Waxman \& Bahcall 1997).
The neutrinos are produced by $\pi^+$ created in interactions
between fireball $\gamma$-rays and
accelerated protons.
The key relation
is between the observed photon energy, $\epsilon_\gamma$,
and the accelerated proton's energy, $\epsilon_p$,
at the photo-meson threshold  of the $\Delta$-resonance.
In the observer's frame,

\begin{equation}
\epsilon_\gamma \, \epsilon_{p} = 0.2 \, {\rm GeV^2} \, \Gamma^2,
\label{eq:keyrelation}
\end{equation}
where phenomenologically the Lorentz factors of the
expanding fireball are $\Gamma > 10^{2}$.
Inserting a typical observed $\gamma$-ray energy of $1$ MeV,
we see that characteristic proton energies
$\sim 2\times10^6$ GeV
are required to produce neutrinos
from pion decay.
Typically, the neutrinos receive $\sim 5$\% of
the proton energy, leading to neutrinos of $\sim 10^{14}$ eV as
stated\footnote{The claim that the spectrum of neutrinos produced
  in interaction with burst (rather than afterglow) photons extends
  to $10^{19}$~eV (\cite{Vietri98}) is due to calculational errors.}.
In the standard picture, these neutrinos result from internal shocks
within the fireball.

In the last two years,
afterglows of GRB's have been discovered in X-ray, optical, and radio
(Costa et al.~1997; van Paradijs et al.~1997; Frail et al.~1997).
These observations confirm (Waxman 1997a; Wijers, Rees, \& M\'esz\'aros
 1997)
standard model predictions (Paczy\'nski \& Rhoads 1993; Katz 1994;
M\'esz\'aros \& Rees 1997; Vietri 1997a)
 of afterglows
that result from the collision of  the expanding fireball with
the surrounding medium. Inserting in Eq.~(\ref{eq:keyrelation}) a
typical afterglow photon energy $\sim 10^2$ eV,
we see that characteristic neutrino energies of order $10^{9}$
GeV may be expected\footnote{Afterglow photons are produced over a wide
range of energies, from radio to X-rays, leading to a broad neutrino
spectrum. As we show in \S4, however, the flux is dominated by neutrinos
in the energy range $10^{17}$--$10^{19}$~eV [cf. Eq. (\ref{eq:Phinu})],
which are
produced in proton interactions with $10$~eV--1~keV photons.}.
$\gamma$-rays of similar energies
are produced by $\pi^0$ decay, but because the fireball
is optically thick at these energies the $\gamma$'s
probably leak out only at much lower energies, $\sim 10$ GeV.

The ultra-high energy neutrinos, $\sim 10^{18}$~eV, are
produced  in the initial stage of the interaction of the
fireball  with its surrounding gas, which  occurs over a time,
$T$,   comparable to the duration of the GRB itself.
Optical--UV photons are radiated  by electrons accelerated in shocks
propagating backward into the
ejecta. Protons are accelerated to high energy
in these ``reverse'' shocks.
The combination of low energy photons and high energy protons produces
ultra-high energy neutrinos via photo-meson interactions, as indicated by
Eq. (\ref{eq:keyrelation}).

Afterglows have been detected in several cases; reverse
shock emission has only been identified for GRB 990123 (Akerlof 1999).
Both the detections and the non-detections are consistent with shocks
occurring with typical model parameters (Sari \& Piran 1999;
M\'esz\'aros \& Rees 1999),
suggesting that reverse shock emission may be common.
The predicted neutrino and gamma-ray emission depends upon parameters
of the  surrounding medium that
can only be estimated once
more observations of the prompt optical afterglow emission are available.

We discuss in \S2 likely plasma conditions in the collisions between
the fireball and the surrounding medium, and in \S3 the
physics of how the ultra-high energy neutrinos are produced. We first
discuss in \S3.1 ultra-high energy cosmic ray (UHECR) production in GRB's
in the light of recent afterglow observations. We show that afterglow
observations provide further support for the model of UHECR production in
GRB's, and address some criticism of the model recently made in the literature
(\cite{GnA99}). Neutrino production is then discussed in \S3.2.
The expected neutrino flux and spectrum is derived in \S4 and the implications
of our results to future experiments are discussed in \S5.

\section{Plasma conditions at the reverse shocks}

We concentrate in this section on the epoch between the time the
expanding fireball first strikes the surrounding medium and
the time when the reverse shocks have erased the memory of the
initial conditions.
After this period,
the expansion approaches the Blandford \& McKee (1976) self-similar
solutions.
The purpose of this section is
to derive the plasma parameters and the UV luminosity and spectrum
expected for typical GRB parameters.
We use the results derived here to estimate in the
following sections
the energy to which protons can be accelerated and to calculate
the production of ultra-high energy neutrinos and
multiple GeV photons.

During self-similar expansion,
the Lorentz factor of plasma at the shock front is
$\Gamma_{BM}=(17E/16\pi n m_p c^2)^{1/2}r^{-3/2}$,
where $E$ is the fireball energy
and $n$ is the surrounding gas number density.
The characteristic time
at which radiation emitted by shocked plasma at radius $r$
is observed by a distant observer is $t\approx r/4\Gamma_{BM}^2 c$
(Waxman 1997b).

The transition to self-similar expansion occurs on a time scale $T$ (measured
in the observer frame)
comparable to the longer of the two time scales set by the initial
conditions: the (observer) GRB duration $t_{\rm GRB}$ and the
(observer) time $T_{\Gamma}$ at which the self-similar Lorentz factor
equals the original ejecta Lorentz factor $\Gamma_i$,
$\Gamma_{BM}(t=T_\Gamma)=\Gamma_i$. Since $t=r/4\Gamma^2_{BM}c$,
\begin{equation}
T=\max\left[t_{\rm GRB}, 6\left({E_{53}\over
n_0}\right)^{1/3}\left({\Gamma_i\over300}\right)^{-8/3}{\rm\, s}\right].
\label{eq:t_tr}
\end{equation}
During the transition,  plasma shocked by the reverse shocks
expands with Lorentz factor close to that given by the self-similar solution,
$\Gamma\simeq\Gamma_{BM}(t=T)$, i.e.
\begin{equation}
\Gamma \simeq 245\left({E_{53}\over n_0}\right)^{1/8}T_1^{-3/8},
\label{eq:Gamma}
\end{equation}
while the unshocked fireball
ejecta propagate with the original expansion Lorentz factor, $\Gamma_i>\Gamma$.
We write Eq.~(\ref{eq:Gamma}) in terms of dimensionless parameters that are
characteristically of order unity in models that successfully
describe observed GRB phenomena. Thus $E=10^{53}E_{53}$~erg,
$T=10 \, T_1$~s, $n=1 \, n_0{\rm cm}^{-3}$, and typically $\Gamma_i
\approx 300$.
Lorentz factors of the reverse shocks in the frames of the
unshocked plasma are mildly relativistic,
$\Gamma_R - 1\sim \Gamma_i  / \Gamma_{\rm BM}$.
If the initial Lorentz factor is extremely large, $\Gamma_i\gg300$,
the transition Lorentz factor
computed from  Eq.~(\ref{eq:t_tr}) and Eq.~(\ref{eq:Gamma})
 remains unchanged, $\Gamma\approx250$,
while the reverse shocks become highly relativistic,
$\Gamma_R\sim\Gamma_i/\Gamma\gg1$.

The observed photon radiation is produced in the fireball model by synchrotron
emission of shock-accelerated electrons. We now summarize the
characteristics of the synchrotron spectrum that leads to
ultra-high energy neutrinos and GeV photons.

Let $\xi_e$ and $\xi_B$ be the fractions of the thermal energy
density $U$ (in the plasma rest frame) that
are carried, respectively,
by electrons and magnetic fields.
The characteristic electron
Lorentz factor (in the plasma rest frame)
is $\gamma_m\simeq \xi_e (\Gamma_R-1)m_p/m_e\simeq
\xi_e(\Gamma_i/\Gamma)m_p/m_e$, where
the thermal energy per proton in the shocked ejecta
is $\simeq (\Gamma_R-1)m_p c^2$. The energy density
$U$ is  $E\approx 4\pi r^2cT
\Gamma^2 U$,
and the number of radiating electrons is $N_e\approx E/\Gamma_i m_p c^2$.
The characteristic(or peak)  energy of
synchrotron photons (in the observer frame) is
\begin{equation}
\epsilon_{\gamma m}^{\rm ob.}
\approx\hbar\Gamma\gamma_m^2{eB\over m_e c}=
0.6\xi_{e,-1}^2\xi_{B,-2}^{1/2}n_0^{1/2}
\left({\Gamma_i\over300}\right)^{2}{\rm\, eV},
\label{eq:e_m}
\end{equation}
and the specific luminosity,
$L_\epsilon=dL/d\epsilon^{\rm ob.}_\gamma$, at
$\epsilon_{\gamma m}^{\rm ob.}$ is
\begin{equation}
L_m\approx
(2\pi\hbar)^{-1}\Gamma{e^3B\over m_e c^2}N_e\approx 
6\times10^{60}\xi_{B,-2}^{1/2}E_{53}^{5/4}T_1^{-3/4}
\left({\Gamma_i\over300}\right)^{-1}n_0^{1/4}{\rm\, s}^{-1},
\label{eq:L_m}
\end{equation}
where $\xi_e=0.1\xi_{e,-1}$, and $\xi_B=0.01\xi_{B,-2}$.
Hereafter, we denote particle energy  in the observer
frame with the super-script ``ob.'', and particle energy measured at the
plasma frame with no super-script (e.g., $\epsilon^{\rm ob.}_{\gamma m}
=\Gamma\epsilon_{\gamma m}$).

Since the reverse shocks are typically
mildly relativistic, electrons are expected to be accelerated in these
shocks to a power law energy distribution,
$dN_e/d\gamma_e\propto\gamma_e^{-p}$ for $\gamma_e>\gamma_m$, with $p\simeq2$
 (Axford, Leer, \& Skadron 1977; Bell 1978; Blandford \& Ostriker 1978). The
specific luminosity extends in this case to energy
$\epsilon_\gamma>\epsilon_{\gamma m}$ as
$L_\epsilon=L_m(\epsilon_\gamma/\epsilon_{\gamma m})^{-1/2}$,
up to photon energy
$\epsilon_{\gamma c}$. Here $\epsilon_{\gamma c}$ is
the characteristic synchrotron frequency of
electrons for which the synchrotron cooling time,
$6\pi m_e c/\sigma_T\gamma_e B^2$, is comparable to the ejecta (rest frame)
expansion time, $\sim r/\Gamma c$. At energy
$\epsilon_\gamma>\epsilon_{\gamma c}$,
\begin{equation}
\epsilon_{\gamma c}^{\rm ob.}\approx
0.3\xi_{B,-2}^{-3/2}n_0^{-1}E_{53}^{-1/2}T_1^{-1/2}
{\rm\, keV},
\label{eq:e_c}
\end{equation}
the spectrum steepens to $L_\epsilon\propto\epsilon_\gamma^{-1}$.

\section{UHECR and neutrino production}

\subsection{UHECR production}

Protons are expected to be accelerated  to high energies in
mildly relativistic shocks within an expanding ultra-relativistic
GRB wind (Waxman 1995a, Vietri 1995). Energies as high as
$\epsilon_p^{\rm ob.}=10^{20}\epsilon_{p,20}^{\rm ob.}$~eV
may be achieved provided the fraction of thermal energy density
carried by magnetic fields, $\xi_B$, is large enough, and provided
shocks occur at large enough radii, so that proton energy loss
by synchrotron emission does not affect acceleration (Waxman 1995a, 1999).
The condition that needs to be satisfied by $\xi_B$,
\begin{equation}
{\xi_B\over\xi_e}>
10^{-2}(\epsilon_{p,20}^{\rm ob.}\Gamma/250)^2L^{-1}_{\gamma,52}\quad,
\label{eq:xi_B}
\end{equation}
where $\Gamma$ is the wind expansion Lorentz factor and
$L_\gamma=10^{52}L_{\gamma,52}{\rm erg\,s}^{-1}$ its $\gamma$-ray luminosity,
is consistent with constraints imposed by afterglow observations.
Afterglow observations imply $\xi_e\sim0.1$ and $\xi_B\sim0.01$
[e.g. Eq. (\ref{eq:e_m})].
The observed distribution of GRB redshifts, which suggests that
most detected GRB's occur at the redshift range
of 1--3 (\cite{KTH98,MnM98,HnF99}), imply that the characteristic
GRB $\gamma$-ray luminosity is $L_\gamma\approx10^{52}{\rm erg\,s}^{-1}$:
For characteristic GRB $\gamma$-ray flux,
$F_\gamma\approx10^{-6}{\rm erg/cm}^2{\rm s}$ in the BATSE
20keV--2MeV range, and adopting the cosmological parameters
$\Omega=0.2$, $\Lambda=0$ and $H_0=75{\rm km/s\,Mpc}$, the luminosity
for a $z=1.5$ burst is $L_\gamma\approx10^{52}{\rm erg\,s}^{-1}$.
This result is consistent with the
more detailed analysis of Mao \& Mo (1998), who obtain a
median GRB luminosity in the 50keV--300keV range (which accounts for $\sim1/3$
of the BATSE range luminosity) of $\approx10^{51}{\rm erg\,s}^{-1}$ for
$\Omega=1$, $\Lambda=0$ and $H_0=100{\rm km/s\,Mpc}$.

The condition that needs to be satisfied to avoid proton synchrotron energy
loss (\cite{W95a}),
\begin{equation}
r>r_{\rm syn}=10^{12}(\Gamma/250)^{-2}(\epsilon_{p,20}^{\rm ob.})^3
\quad{\rm cm},
\label{eq:r_syn}
\end{equation}
is clearly satisfied in the present context, as
reverse shocks arise at $r\approx4\Gamma^2cT\approx10^{17}{\rm cm}
\gg10^{12}$~cm. Thus, synchrotron losses of protons accelerated to high energy
at the radii where reverse shocks are expected to arise
are negligible.

We note that it has recently been claimed (Gallant \& Achterberg 1999)
  that acceleration of
  protons to $\sim10^{20}$~eV in the {\it highly-relativistic external}
  shock driven by the fireball into its surround medium is impossible.
  Regardless of whether this claim is correct or not, it is not relevant to
  the model proposed in Waxman (1995a) and discussed here, in which protons are
  accelerated in the {\it mildly-relativistic internal (reverse)} shocks.

Finally, improved constraints from afterglow observations on the energy
generation rate of GRB's provide further support to the GRB model
of UHECR production. For an open universe, $\Omega=0.2$, $\Lambda=0$ and
$H_0=75{\rm km/s\,Mpc}$, the GRB rate per unit volume required to account for
the observed BATSE rate is $R_{\rm GRB}\approx10^{-8}{\rm Mpc^{-3}yr^{-1}}$,
assuming a
constant comoving GRB rate. Present data does not allow to distinguish
between models in which the GRB rate is evolving with redshift, e.g. following
star formation rate, and models in which it is not evolving, since in
both cases most detected GRB's occur at the redshfit range of 1--3
(\cite{HnF99}). Thus, $R_{\rm GRB}$ provides a robust estimate of the rate
at $z\sim1$, while the present, $z=0$, rate may be lower by a factor of
$\sim8$ if strong redshift evolution is assumed. This implies that the present
rate of $\gamma$-ray energy generation by GRB's is in the range of
$10^{44}{\rm erg/Mpc^{3}yr}$ to $10^{45}{\rm erg/Mpc^{3}yr}$, remarkably
similar to the energy generation rate required to account for the observed
UHECR flux above $10^{19}$~eV, $\sim10^{44}{\rm erg/Mpc^{3}yr}$
(\cite{W95b,WnB99}).

\subsection{Neutrino production}

The photon distribution in the wind rest frame is isotropic. Denoting
by $n_\gamma(\epsilon_\gamma)d\epsilon_\gamma$ the
number density
(in the wind rest frame) of photons in the energy range $\epsilon_\gamma$
to $\epsilon_\gamma+d\epsilon_\gamma$,
the fractional energy loss rate of
a proton with energy $\epsilon_p$ due to pion production is
\begin{eqnarray}
t_\pi^{-1}(\epsilon_p)\equiv&&
-{1\over\epsilon_p}{d\epsilon_p\over dt}\nonumber\\=&&
{1\over2\gamma_p^2}c\int_{\epsilon_0}^\infty
d\epsilon\,\sigma_\pi(\epsilon)
\xi(\epsilon)\epsilon\,\int_{\epsilon/2\gamma_p}^\infty dx\,
x^{-2}n(x)\,,
\label{eq:pirate}
\end{eqnarray}
where $\gamma_p=\epsilon_p/m_pc^2$, $\sigma_\pi(\epsilon)$ is the cross
section for pion production for a photon with energy $\epsilon$ in the
proton rest frame, $\xi(\epsilon)$ is the average fraction of energy
lost to the pion, and $\epsilon_0=0.15 \, {\rm GeV}$ is the threshold
energy.

The photon density is related to the observed luminosity by
$n(x)=L_\epsilon(\Gamma x)/(4\pi r^2c\Gamma x)$.
For proton Lorentz factor
$\epsilon_0/2\epsilon_{\gamma c}\ll\gamma_p<\epsilon_0/2\epsilon_{\gamma m}$,
photo-meson production is dominated by interaction with photons
in the energy range
$\epsilon_{\gamma m}<\epsilon_\gamma\ll\epsilon_{\gamma c}$,
where $L_\epsilon\propto\epsilon_\gamma^{-1/2}$. For this photon spectrum,
the contribution to the first integral of Eq. (\ref{eq:pirate}) from
photons at the $\Delta$ resonance is comparable to that of photons of
higher energy, and we obtain
\begin{equation}
t_\pi^{-1}(\epsilon_p)\simeq {2^{5/2}\over2.5}
{L_m\over4\pi r^2\Gamma}
\left({\epsilon_{\rm peak}\over \gamma_p \epsilon_{\gamma
m}}\right)^{-1/2}
 {\sigma_{\rm peak}\xi_{\rm peak}
\Delta\epsilon\over\epsilon_{\rm peak}}.
\label{eq:taupi}
\end{equation}
Here, $\sigma_{\rm peak}\simeq5\times10^{-28}{\rm cm}^2$ and
$\xi_{\rm peak}\simeq0.2$
at the resonance $\epsilon=\epsilon_{\rm peak}=0.3 \, {\rm GeV}$,
and $\Delta\epsilon\simeq0.2 \, {\rm GeV}$ is the peak width.
The time available
for proton energy loss by pion production is comparable to
the expansion time as measured in the wind rest frame,
$\sim r/\Gamma c$.
Thus, the fraction of energy lost by protons to pions is
\begin{eqnarray}
f_\pi(\epsilon_p^{\rm ob.})\approx && 0.05
\left({L_m\over6\times10^{60}{\rm s}^{-1}}\right)
\left({\Gamma\over250}\right)^{-5} T_1^{-1}\nonumber\\ &&\times
(\epsilon_{\gamma m,{\rm eV}}^{\rm ob.}\epsilon_{p,20}^{\rm ob.})^{1/2}.
\label{eq:fpi}
\end{eqnarray}
Eq. (\ref{eq:fpi}) is valid for protons in the energy range
\begin{eqnarray}
4\times10^{18}\left({\Gamma\over250}\right)^2
(\epsilon_{\gamma c,{\rm keV}}^{\rm ob.})^{-1}{\rm eV}
<\epsilon_{p}^{\rm ob.}<\nonumber\\
4\times10^{21}\left({\Gamma\over250}\right)^2
(\epsilon_{\gamma m,{\rm eV}}^{\rm ob.})^{-1}{\rm eV}
\,.
\label{eq:ep_range}
\end{eqnarray}
Such protons interact with photons in the energy range $\epsilon_{\gamma m}$
to $\epsilon_{\gamma c}$, where the photon spectrum
$L_\epsilon\propto\epsilon_{\gamma c}^{1/2}$ and the number of photons above
interaction threshold is $\propto\epsilon_p^{1/2}$.
At lower energy, protons interact with photons of energy
$\epsilon_\gamma>\epsilon_{\gamma c}$, where
$L_\epsilon\propto\epsilon^{-1}$ rather then
$L_\epsilon\propto\epsilon^{-1/2}$. At these energies therefore
$f_\pi\propto\epsilon_p^{\rm ob.}$.

Since the flow is ultra-relativistic, the results given above
are independent
of whether the wind is spherically symmetric or jet-like, provided the jet
opening angle is $>1/\Gamma$ .
For a jet-like wind, $L_m$ is the luminosity that
would have been produced by the wind if it were spherically symmetric.

\section{Neutrino spectrum and flux}

Approximately
 half of the energy lost by protons goes into $\pi^0$~'s and the
other half to $\pi^+$~'s.  Neutrinos are produced by the decay of $\pi^+$'s,
$\pi^+\rightarrow\mu^++\nu_\mu
\rightarrow e^++\nu_e+\overline\nu_\mu+\nu_\mu$.
The mean pion energy is $20\%$
of the proton energy. This energy is roughly
evenly distributed between
the $\pi^+$ decay products. Thus, approximately half
the energy lost by protons of energy $\epsilon_p$ is converted to neutrinos
with energy $\sim0.05\epsilon_p$. Eq. (\ref{eq:ep_range}) implies that
the spectrum of neutrinos below
$\epsilon_{\nu b}^{\rm ob.}\approx10^{17}(\Gamma/250)^2
(\epsilon_{\gamma c,{\rm keV}}^{\rm ob.})^{-1}{\rm eV}$ is harder
by one power of the energy then the proton spectrum, and
by half a power of the energy at higher energy. For a power law
differential spectrum of accelerated protons $n(\epsilon_p)\propto
\epsilon_p^{-2}$, as expected for Fermi acceleration and which
could produce the observed spectrum of ultra-high energy cosmic rays
 (Waxman 1995b), the differential neutrino spectrum is $n(\epsilon_\nu)
\propto\epsilon_\nu^{-\alpha}$ with $\alpha=1$ below the break and
$\alpha=3/2$ above the break.

The energy production rate required to produce the observed
flux of ultra-high energy cosmic-rays, assuming that the sources are
cosmologically distributed, is (Waxman 1995b)
\begin{equation}
E^2_{CR}d\dot N_{CR}/dE_{CR}\approx10^{44}{\rm erg\ Mpc}^{-3}{\rm yr}^{-1}.
\label{eq:E_CR}
\end{equation}
If GRB's are indeed the sources of ultra-high energy cosmic rays,
then Eqs. (\ref{eq:fpi},\ref{eq:ep_range})
imply that the expected neutrino intensity is
\begin{equation}
\epsilon_\nu^2\Phi_\nu\approx 10^{-10}{f_\pi^{[19]}\over0.1}
\left({\epsilon_\nu^{\rm ob.}\over10^{17}{\rm eV}}\right)^{\beta}
{\rm GeV\,cm}^{-2}{\rm s}^{-1}{\rm sr}^{-1},
\label{eq:Phinu}
\end{equation}
where $f_\pi^{[19]}\equiv f_\pi(\epsilon_{p,20}^{\rm ob.}=2)$ and
$\beta=1/2$ for $\epsilon_\nu^{\rm ob.}>10^{17}{\rm eV}$
and $\beta=1$ for $\epsilon_\nu^{\rm ob.}<10^{17}{\rm eV}$.
The fluxes of all neutrinos are similar and given by Eq. (\ref{eq:Phinu}),
$\Phi_{\nu_\mu}\approx\Phi_{\bar\nu_\mu}\approx\Phi_{\nu_e}\approx\Phi_\nu$.
Eq. (\ref{eq:Phinu}) is obtained by integrating the neutrino generation
rate implied by Eq. (\ref{eq:E_CR}) and (\ref{eq:fpi}) over cosmological
time, under the assumption that the generation rate is independent of
cosmic time (Waxman \& Bahcall 1999).
If the GRB energy generation rate increases with redshift in a manner
similar to the evolution of the QSO luminosity density, which exhibits
the fastest known redshift evolution, the expected neutrino flux would
be $\sim3$ times that given in Eq. (\ref{eq:Phinu}) (Waxman \& Bahcall
1999).

The neutrino flux is expected to be strongly suppressed at energy
$\epsilon^{\rm ob.}_\nu>10^{19}$~eV, since protons are not expected to be
accelerated to energy $\epsilon^{\rm ob.}_p\gg10^{20}$~eV.
If protons are accelerated to much higher energy, the $\nu_\mu$
flux may extend to $\sim10^{21}$~eV. At higher energy
the ejecta expansion time $\Gamma T$ is shorter than the pion decay time,
leading to strong suppression of $\nu_\mu$ flux due to adiabatic energy loss
at $\epsilon^{\rm ob.}_\nu>10^{21}T_1(\Gamma/250)^2$~eV.
Adiabatic energy loss of muons will
suppress the $\bar\nu_\mu$ and $\nu_e$ flux at
$\epsilon^{\rm ob.}_\nu>10^{19}T_1(\Gamma/250)^2$~eV.

\section{Discussion}

When the expanding fireball of a GRB collides with the surrounding medium,
reverse shocks are created that give rise to observed afterglow
by synchrotron radiation.  The specific luminosity and energy spectrum
describing these processes are given in Eq.~(\ref{eq:e_m})-Eq.~(\ref{eq:e_c}).
For typical values of the plasma parameters and fireball Lorentz factor,
these equations are consistent with the afterglow observations.
The burst GRB 990123 was especially luminous
and also for this burst the relations Eq.~(\ref{eq:e_m})-Eq.~(\ref{eq:e_c})
are consistent with the observation of a reverse shock and the
other afterglow phenomena.

If protons are accelerated in GRB's up
to energies $\sim 10^{20}$ eV, then the expected flux of high energy neutrinos
is given by Eq.~(\ref{eq:Phinu}) and (\ref{eq:fpi}).
Muon energy loss suppresses the $\bar\nu_\mu$ and $\nu_e$ flux
above $\sim10^{19}$~eV, while pion energy loss suppresses the
$\nu_\mu$ flux only above $\sim10^{21}$~eV. Since protons are not
expected to be accelerated to  $\gg10^{20}$ eV (Waxman 1995a), the
energy beyond which the $\nu_\mu$ flux is suppressed will likely be determined
by the maximum energy of accelerated protons. Measuring the maximum
neutrino energy will set a lower limit to the maximum proton energy.
The predicted flux is sensitive to the value of the Lorentz factor
of the reverse shock (see Eq.~\ref{eq:fpi}), but this value is given robustly
by Eq.~(\ref{eq:Gamma}) as $\Gamma \simeq 250$.

Will the ultra-high energy neutrinos predicted in this letter be detectable?
The sensitivities of high-energy neutrino detectors
have not been determined for
ultra-high energy neutrinos whose time of occurrence is known to
within $\sim 10$ s and whose direction on the sky is known accurately.
Special techniques may enhance the detection of GRB neutrinos (see below).

Planned $1{\rm km}^3$
detectors of high energy neutrinos include
ICECUBE, ANTARES, NESTOR (Halzen 1999) and
NuBE (Roy, Crawford, \& Trattner 1999).
Neutrinos are detected by observing optical Cherenkov light emitted by
neutrino-induced muons.
The probability $P_{\nu\mu}$ that a neutrino
would produce a high energy muon with the currently
required long path within the detector is
$P_{\nu\mu}\approx3\times10^{-3}(\epsilon_\nu/10^{17}{\rm eV})^{1/2}$
(Gaisser, Halzen \& Stanev 1995; Gandhi, Quig, Reno, \& Sarcevic 1998).
Using (\ref{eq:Phinu}), the expected detection rate of muon neutrinos
is $\sim0.06/{\rm km}^2{\rm yr}$ (over $2\pi$~sr), or $\sim 3$ times
larger if GRB's evolve like quasars.
GRB neutrinos may be detectable in these experiments because the knowledge
of neutrino direction and arrival time may relax the requirement for long
muon path within the detector.

Air-showers could be used to detect ultra-high energy neutrinos.
The neutrino acceptance of the planned Auger detector,
$\sim10^4{\rm km^3}{\rm sr}$ (Parente \& Zas 1996), seems too low.
The effective area of proposed space detectors (Linsley 1985;
Takahashi 1995)
may exceed $\sim10^6{\rm km^2}$
at $\epsilon_\nu>2\times10^{19}$~eV, detecting several tens of
GRB correlated events per year, provided that the neutrino flux extends to
$\epsilon_\nu>2\times10^{19}$~eV. Since, however, the GRB neutrino flux
is not expected to extend well above  $\epsilon_\nu\sim10^{19}$~eV, and since
the acceptance of space detectors decrease rapidly below
$\sim10^{19}$~eV, the detection rate of space detectors would depend
sensitively on their low energy threshold.

As explained in Waxman \& Bahcall (1997), $\nu_\tau$'s are not expected to be
produced in the GRB. However, the strong mixing between $\nu_\mu$ and
$\nu_\tau$ favored by Super-Kamiokande observations of atmospheric
neutrinos indicates that the flux of $\nu_\mu$ and $\nu_\tau$
should be equal at Earth. This conclusion would not hold if the less
favored alternative of $\nu_\mu$ to $\nu_{\rm sterile}$ occurs.

The decay of $\pi^0$'s produced in photo-meson interactions would lead
to the production of $\sim10^{19}$~eV photons. For the photon luminosity
and spectrum given in Eqs. (\ref{eq:e_m}--\ref{eq:e_c}), the fireball
optical depth for pair production is higher than unity for
$\epsilon_\gamma>10$~GeV. Thus, the ultra-high energy photons would
be degraded and will escape the
fireball as multi-GeV photons. Since in order for GRB's to be the sources
of ultra-high energy protons similar energy should be produced
in $\sim1$~MeV photons and $\sim10^{20}$~eV protons
 (Waxman 1995b), the expected multi-GeV
integrated luminosity is $\approx10\%$ of the $\sim1$~MeV integrated
luminosity, i.e. $\sim10^{-6}{\rm erg/cm}^2$. Such multi-GeV emission has been
detected in several GRB's on time scale of $>10$~s following the GRB,
and may be common (Dingus 1995). This is not,
however, conclusive evidence for proton acceleration to ultra-high
energy. For the parameters adopted in this paper, inverse-Compton
scattering of synchrotron photons may also produce the observed
multi-GeV photons. More
wide spectrum observations, of optical to $>10$~GeV photons,
are required to determine
whether the observed multi-GeV emission on $\sim10$~s
is due to inverse-Compton or $\pi^0$ decay.

We note here that Multi-GeV photon production by
  synchrotron emission of $\sim10^{20}$~eV protons accelerated at the
  {\it highly-relativistic external} shock driven by the fireball into its
  surround medium has been discussed in Vietri (1997b) and Bottcehr \&
Dermer (1998). However,
  protons are not likely to be accelerated to such energy at the external
  shock (Gallant \& Achterberg 1999), and, moreover, even if acceleration is
possible, the
  fraction of proton energy lost by synchrotron emission at the
radii where external shocks occur is $\ll1$, [see
  discussion following Eq. (\ref{eq:r_syn})] and hence the expected flux is
  much smaller than the inverse-Compton or $\pi^0$ decay flux.

\acknowledgments
JNB was supported in part by NSF PHY95-13835. EW was supported in part
by  BSF Grant 9800343, AEC Grant 38/99 and MINERVA Grant.

\end{document}